\documentstyle[12pt]{article}
\begin{document}
\def\th{\theta}
\def\de{\Delta}
\def\cj{{\Im}}
\def\kb{k_B}
\def\vk{{\bf k}} 
\def\vq{{\bf Q}} 
\def\vri{{\bf R}_i} 
\def\vrj{{\bf R}_j} 
\def\lsim{\stackrel{\textstyle <}{\sim}} 
\def\beq{\begin{equation}}
\def\enq{\end{equation}}
\def\beqn{\begin{eqnarray}}
\def\eeqn{\end{eqnarray}}
\def\pl{\parallel}
\baselineskip19pt

\begin{flushright}
MRI-PHY/96-38 \\ 
\end{flushright}

\begin{center}
{\large{\bf Phase Diagram of the Half-Filled  Extended Hubbard Model 
            in Two Dimensions}}\\
\vspace{0.8cm} 
{\bf Biplab Chattopadhyay$^\star$ and D. M. Gaitonde$^\dagger$}\\
{The Mehta Research Institute of Mathematics \& Mathematical Physics,}\\  
{10 Kasturba Gandhi Marg, Allahabad 211002, INDIA} 
\end{center} 
\vspace{1cm}

\begin{abstract}
We consider an extended Hubbard model of interacting fermions on 
a lattice. The fermion kinetic energy corresponds to a tight binding
Hamiltonian with nearest neighbour ($-t$) and next nearest
neighbour ($t^{\prime}$) hopping matrix elements.
In addition to the onsite Hubbard interaction ($U$)
we also consider a nearest neighbour repulsion ($V$).
We obtain the zero temperature phase diagram of our model
within the Hartree-Fock approximation.
 We consider 
ground states having charge and spin density wave ordering as well 
as states with orbital antiferromagnetism or spin nematic order. 
The latter two states correspond to particle-hole binding with
$d_{x^2-y^2}$ symmetry in the charge and spin channels respectively.
For 
$t^\prime = 0$, only the charge density wave and spin density wave states 
are energetically stable. For non-zero $t^\prime$, we find that orbital 
antiferromagnetism (or spin nematic) order is stable over a finite portion 
of the phase diagram at weak coupling. This region of stability is seen to 
grow with increasing values of $t^\prime$. 
\end{abstract}
\vspace{1ex} 

\vspace{1ex} 
\noindent To appear in {\bf Phys. Rev. B (Brief Reports)}

\vfill
\noindent \rule{10cm}{0.1mm}\\
{\small $^\star$Present email: biplab@hp1.saha.ernet.in,\,\,$^\dagger$email: dattu@mri.ernet.in\,;}\\

\newpage 

The phase diagram of the 2-d Hubbard model has been a subject of intense 
study over the past few years, ever since the discovery of the high 
temperature superconductors. The strange properties of these materials 
are believed to arise because of strong electronic correlation and the 
2-d Hubbard model or its strong coupling limit, the t-J model, have 
frequently been used as a starting point for theoretical attempts 
at understanding these phenomena. Even earlier, this model has been 
studied as the minimal model necessary for describing a host of 
phenomena ranging from metal-insulator transitions in the transition 
metal oxides to magnetism in itinerant electron systems. 

On a square lattice, at half-filling, a tight-binding Hamiltonian with 
hopping restricted to nearest neighbours, has a Fermi surface that is 
perfectly nested, so that at zero temperature the ground state of the 
system is a two sublattice spin density wave (SDW) state for arbitrarily 
small onsite repulsion (U). 
Detailed studies \cite{schr} of the SDW Hartree-Fock (HF) solutions 
as well as a random-phase-approximation (RPA) study of the collective excitation 
spectrum have shown that this approach is effective even at large 
values of U where the magnetization and spin wave dispersion obtained 
from a mapping to the Heisenberg model are recovered. The introduction 
of a nearest-neighbour repulsion (V) introduces a tendency towards charge 
ordering in the form of a charge density wave (CDW) and for sufficiently 
large values of V ($V>U/4$), the CDW state becomes energetically  
preferred to the SDW state. 

It has been known since the early work of Halperin and Rice \cite{halp}
that there exist forms of particle-hole ordering other than the CDW and 
SDW state. In this context, some other ground states which have been 
considered are orbital antiferromagnetism (OAF) and spin nematic (SN) 
[3-6]. These ground states can be thought of as arising from d-wave 
pairing in the particle-hole charge and spin channels respectively 
\cite{nayak}. The OAF and SN states have charge and spin currents in 
every elementary plaquette of the square lattice with neighbouring 
plaquettes having their currents oriented in opposite directions. 
These states are thus similar to the staggered flux phases which were 
studied earlier \cite{affleck} as possible ground states of the t-J 
model, in the strong correlation limit. 

In this paper we present a Hartree-Fock study of the zero 
temperature phase diagram of the extended Hubbard model at
half-filling on a square lattice especially focussing on the
competition of the CDW and SDW states with the OAF or SN states.
Our model has a kinetic energy corresponding to
nearest-neighbour and next nearest-neighbour hopping whose
amplitudes are $t$ and $t^\prime$ respectively and an onsite (U)
and nearest neighbour (V) repulsive interactions. For $t^\prime = 0$
we find that the CDW and SDW states are the only stable ground
states of the system and are separated by a phase boundary at $V=U/4$. 
The SDW and CDW order parameters
do not vanish at the phase boundary and the transition is
discontinuous. However, for finite values of $t^\prime$ the
OAF and SN states which are energetically degenerate within
the HF approximation, are found to be lower in energy than
the CDW and SDW states over a finite portion of the phase
diagram, whose size grows with increasing $t^\prime$. 

Our main findings (summarized in Fig.2) indicate that the absence 
of nesting (which is destroyed by a finite $t^\prime$) favours the 
occurence of the OAF and SN states. This is significant since most 
real materials have non-nested Fermi surfaces. Stable OAF and SN 
states are found to occur in the weak correlation limit 
($U,V$ less than the bandwidth) where the HF approximation
is expected to be reliable. The d-wave character of the 
particle-hole binding involved in the formation of the OAF and SN states, 
is interesting in the light of the observation of a pseudo-gap
with a d-wave symmetry in photoemission experiments
\cite{marsh,ding} in the underdoped cuprates which has been
interpreted variously as evidence for a paired state where 
superconductivity has been destroyed by phase fluctuations 
\cite{emery} or a manifestation of the spinon dispersion
predicted by slave boson theories of the t-J model \cite{laugh}. 
It is worth pointing out that d-wave pairing in the
particle-hole channel, such as occurs in the OAF and SN phases,
also leads to a similar one-particle excitation spectrum which is 
qualitatively consistent with the photoemission data. 

We consider the extended Hubbard model on a square lattice, whose
Hamiltonian is given by
\beq 
 H = \sum_{i,j,\sigma} t_{ij}{c_{i,\sigma}^\dagger}{c_{j,\sigma}}
       - \mu \sum_{i,\sigma}\hat{n}_{i,\sigma} 
       + U \sum_{i} \hat{n}_{i,\uparrow} \hat{n}_{i,\downarrow}
       + {V\over 2} \sum_{i,\sigma,\sigma^\prime} \sum_{\delta_{nn}}
    \hat{n}_{i,\sigma} 
    \hat{n}_{i+\delta_{nn},\sigma^\prime}   
\enq 
The hopping matrix elements $t_{ij}$ are chosen to be $t_{ij} =-t$ 
for $i,\,j$ nearest-neighbours, $t_{ij}=t^\prime$ for $i,\,j$ 
next nearest neighbours and zero otherwise. Here $\mu$ is the 
chemical potential and $U$ and 
$V$ are the onsite and nearest neighbour repulsive interactions 
respectively. We work throughout at half-filling with one electron
per site on the average. We consider mean-field states characterized by
an anomalous non-zero ground state expectation value of the average 
$\langle c_{\vk,\alpha}^\dagger c_{\vk+\vq, \beta }\rangle$ with  
$\vq = (\pi,\pi)$ being the nesting vector of the non-interacting
Fermi surface for $t^\prime = 0$.  

The CDW order parameter results from a real space decoupling of the form 
$$
\left\langle\hat{n}_{i, \sigma}\right\rangle = {1\over 2}
                 + {p\over 2} \cos(\vq.\vri)   \eqno(2a)
$$
whereas the SDW state (we consider a z-polarized state) is described by 
$$
\left\langle\hat{n}_{i, \sigma}\right\rangle = {1\over 2}
                 + \sigma{m\over 2} \cos(\vq.\vri)   \eqno(2b)
$$
$\sigma=+(-) 1$\,\, for up (down) spins. These order parameters describe a 
periodic
modulation in the charge and spin density respectively resulting in 
a doubling of the unit cell. $p$ and $m$ are the amplitudes
of the induced density modulation. 
\addtocounter{equation}{1}

The OAF  state on the other hand has non-zero intersite averages
of the form
$$
\left\langle c_{i, \sigma}^\dagger c_{i\pm\hat{x} \sigma}\right\rangle
                  = \imath g\, \cos(\vq.\vri)\eqno(3a)  
$$ 
and 
$$
\left\langle c_{i, \sigma}^\dagger c_{i\pm\hat{y} \sigma}\right\rangle
                  = - \imath g\, \cos(\vq.\vri)\eqno(3b) 
$$ 
The SN state (we consider a z-polarised state) has
$$
\left\langle c_{i, \sigma}^\dagger c_{i\pm\hat{x} \sigma}\right\rangle
                  = \imath \ell\sigma\, \cos(\vq.\vri)\eqno(4a)  
$$ 
and 
$$
\left\langle c_{i, \sigma}^\dagger c_{i\pm\hat{y} \sigma}\right\rangle
                  = - \imath \ell\sigma\, \cos(\vq.\vri)\eqno(4b) 
$$ 
It is easy to see from Eqns.(3) and (4) that the OAF and SN states
have charge and spin currents (whose magnitude is proportional to $g$
and $l$ respectively) moving along the elementary square
plaquettes of the lattice with neighbouring plaquettes having their
currents in opposite directions (see Fig.1). These states 
have been discussed earlier by several workers [3-6].
However, their stability with respect to the CDW and SDW states has
never been precisely established. Earlier calculations [3-5] have
approximately evaluated the appropriate  RPA susceptibilities
in the normal state. 
The divergence of these susceptibilities
gives the mean field transition and the temperature
where this happens is the mean field transition temperature.
It is well-known \cite{mermin} 
that in two dimensions, enhanced fluctuations
prevent long range order from developing at any finite
temperature. The RPA fails to take proper
account of these enhanced fluctuations and thus predicts
 a spurious finite temperature phase transition. 
The  system continues to be a Fermi liquid
below the mean field transition temperature.
However, this temperature
is  a crossover scale below which the Fermi liquid begins
to develop short range correlations  appropriate
to the type of broken symmetry considered.
The corresponding correlation length increases as the temperature
is reduced and diverges at $T=0$.
A comparision of the mean field transition temperatures of the different
states is a rough indication of their relative stability.
However, an accurate determination of the zero temperature
phase diagram of the model Hamiltonian requires a complete solution
of the self-consistent equations (obtained by minimizing the ground 
state energy)
for the chemical potential and the order parameter.
A comparision of the corresponding ground state energies then
fixes the stable ground state for the parameters considered.
The resulting phase diagram is the main result of this paper.

\addtocounter{equation}{2}

We first obtain the self-consistent equations for the order parameter
and the chemical potential (which has to be determined self-consistently 
for $t^\prime \neq 0$). On substituting the mean field ansatz of Eqns(2-4)
in the Hamiltonian of Eqn(1) we find that the mean field Hamiltonian, for the various states
considered, is given by 
\beqn 
 H_{MF}^s &=& \sum_{\vk,\alpha}^{RBZ} \left[(\xi_\vk -\mu_)
            {c_{\vk,\alpha}^\dagger}{c_{\vk,\alpha}} + 
            (\xi_{\vk+\vq} -\mu_){c_{{\vk+\vq},\alpha}^\dagger}
            {c_{{\vk+\vq},\alpha}}\right]  \nonumber\\ 
          & & + \sum_{\vk,\alpha,\beta}^{RBZ}
            \left[(\Delta_{\alpha,\beta}^s(\vk)
            {c_{\vk,\alpha}^\dagger}{c_{{\vk+\vq},\beta}} + {\rm h.c} \right] 
           + X_s 
\eeqn  
The reduced Brillouin zone (RBZ) is formed due to the modulation 
with wavevector $\vq$, which introduces a two-sublattice structure resulting 
in a folding up of the original Brillouin zone. The tight-binding dispersion 
is $\xi_{\vk} = -2t[\cos k_x + \cos k_y] + 4t^\prime \cos k_x\,\cos k_y$ 
and $\Delta_{\alpha,\beta}^s(\vk)$ is the generalized order parameter where 
$s=1,2,3,\,\&\, 4$ correspond to CDW, SDW, OAF and SN states respectively. 
Thus, 
$$
~~~~~~~~~~~~~~~\Delta_{\alpha,\beta}^1(\vk) = \delta_{\alpha,\beta} ({U\over 2} - 4V)p
                    \hfill~~~~~~~~~~~~~~~~~~~~~~~~~~~~~~~~~~~{\rm for~the~CDW ~state} \eqno(6a) 
$$ 
$$
~~~~~~~~~~~~~~\Delta_{\alpha,\beta}^2(\vk) = \sigma_{\alpha,\beta}^z (-{U\over 2})m 
                   \hfill~~~~~~~~~~~~~~~~~~~~~~{\rm for~the ~z-polarized ~SDW ~state} \eqno(6b) 
$$ 
$$
~~~~~~~~~~~~~\Delta_{\alpha,\beta}^3(\vk) = \imath \delta_{\alpha,\beta}
                               (\cos k_x - \cos k_y)(-2V)g 
                   \hfill~~~~~~~~~~~~~~~~~~~{\rm for~ the~ OAF~ state} \eqno(6c) 
$$ 
$$
{\rm and}~~~~~~~~\Delta_{\alpha,\beta}^4(\vk) = \imath \sigma_{\alpha,\beta}^z 
                               (\cos k_x - \cos k_y)(-2V)\ell  
                    \hfill~~~~{\rm for~ the~ z-polarized~ SN ~state} \eqno(6d) 
$$ 
where $\sigma^z$ is a Pauli matrix. In writing Eqn(5) we have ignored 
the Hartree contribution $(U/2 + 4V)$ which shifts the energy of each state 
by an equal amount and can therefore be absorbed in the chemical 
potential. The constant piece $X_s$ in the respective ground state 
energies is given by 
$$
X_1 = (2V - {U\over 4}) p^2 N \eqno(7a)
$$
$$
X_2 =  {U\over 4} m^2 N ~~~~~~~~~ \eqno(7b)
$$
$$
X_3 = 4V g^2 N ~~~~~~~~~ \eqno(7c)
$$
$$
X_4 = 4V \ell^2 N ~~~~~~~~~~ \eqno(7d)
$$
where N is the total number of sites. In writing Eqn(7) we have once 
again ignored the Hartree contribution of $-(U/4 + 2V)N$ which merely 
shifts the ground-state energies of each of the mean field solutions 
by an equal amount. 
\addtocounter{equation}{2}

The mean field Hamiltonian of Eqn(5) is quadratic and can be
diagonalized by a canonical transformation. The folding of the Brillouin
zone, caused by the ($\pi, \pi$) modulation, results in the formation
of two bands whose dispersion is given by

$$
e_s^{(+)}(\vk) = (\eta_\vk - \mu) + \sqrt{\epsilon_\vk^2 + \Delta_s^2(\vk)}\eqno(8a) 
$$ 
and 
$$
e_s^{(-)}(\vk) = (\eta_\vk - \mu) - \sqrt{\epsilon_\vk^2 + \Delta_s^2(\vk)}\eqno(8b) 
$$ 
Here $\Delta_1(\vk) = (U/2 - 4V)p$, $\Delta_2(\vk) = -Um/2$, 
$\Delta_3(\vk) = - 2Vg(\cos k_x - \cos k_y)$ and 
$\Delta_4(\vk) = - 2V\ell (\cos k_x - \cos k_y)$, 
$\eta_\vk = (\xi_\vk + \xi_{\vk+\vq})/2$,  
$\epsilon_\vk = (\xi_\vk - \xi_{\vk+\vq})/2$ and $\xi_\vk$ is the  
band dispersion in the non interacting limit. 
For $t^\prime = 0$, nesting of the Fermi surface ensures that
$\eta_\vk=0$. Then, the chemical potential stays pinned to it's
non-interacting value ($\mu=0$) and the lower band ($e_\vk^{(-)}$) 
is completely full whereas the upper band is completely empty. 
However when $t^\prime$ becomes finite, $\eta_\vk\neq0$ and 
the bands begin to overlap. It is easy to see that CDW 
($\Delta_1(\vk)$) and SDW ($\Delta_2(\vk)$) states are isotropic 
(s-wave) whereas the OAF and SN states ($\Delta_3(\vk)$ and 
$\Delta_4(\vk)$)) states have order parameters of $d_{x^2-y^2}$ 
character. 

Using the diagonalized form of the Hamiltonian, we obtain the 
self-consistent equations for the chemical potential $\mu$ and 
the order parameters, which are given by
$$
{1\over 2} = {1\over N} \sum_\vk^{RBZ}
                 \left[\Theta\!\!\left(-e_s^{(-)}(\vk)\right)
                  + \Theta\!\!\left(-e_s^{(+)}(\vk)\right)\right] \eqno(9a)
$$ 
and 
$$
{1\over{\Gamma_s}} = {1\over N} \sum_\vk^{RBZ}
                 {w_\vk^s\,{\left[\Theta\!\!\left(-e_s^{(-)}(\vk)\right)
                  + \Theta\!\!\left(-e_s^{(+)}(\vk)\right)\right]}\over
                     {\sqrt{\epsilon_\vk^2 + \Delta_s^2(\vk)}}} \eqno(9b)
$$ 
Here $\Gamma_1 = 4V-U/2$, $\Gamma_2 = U/2$, $\Gamma_3=\Gamma_4 = V/2$ and 
$w_\vk^s$ is the symmetry factor of the order parameters with 
$w_\vk^1 = w_\vk^2 = 1$ (for SDW \& CDW) and 
$w_\vk^3 = w_\vk^4 = (\cos k_x - \cos k_y)^2$ (for OAF \& SN). 
\addtocounter{equation}{2}

We have solved Eqn(9) numerically to determine the self-consistent 
values of the chemical potential $\mu$ and the gap function 
$\Delta_s(\vk)$. Self-consistent solutions for all the four mean field 
states exist for almost all values of $U/t$ and $V/t$ and therefore it 
becomes necessary to compare the ground state energies in order to 
decide the relative stabilities of these states. The ground state 
energy of a given state can be easily found from the 
diagonalized Hamiltonian and is given by 
\beq 
E^s = {X_s} + {2\over N} \sum_k^{RBZ} 
      \left[ e_s^{(-)}(\vk)\, \Theta\!\!\left(-e_s^{(-)}(\vk)\right) 
      + e_s^{(+)}(\vk)\, \Theta\!\!\left(-e_s^{(+)}(\vk)\right)\right] 
\enq 

It is easy to see from a comparison of Eqns(9) and (10) that the 
OAF and SN states are degenerate within the Hartree-Fock approximation 
for our model. Further, the dispersion of the energy bands 
($e^{(-)}(\vk)$ and $e^{(+)}(\vk)$) is also identical.   
These states are very different physically. The OAF state has a 
staggered orbital magnetic moment produced by the circulating 
charge currents which should be observable in  magnetic 
neutron scattering  experiments,
whereas the spin-currents present in the SN 
state would not directly couple to the neutron magnetic moment. 
However, they are identical in their one-particle properties as 
far as experiments which probe their band structure are concerned. 

We use the self-consistent values of $\Delta_s(\vk)$ and $\mu$, 
obtained from the solution of Eqn(9), to evaluate the ground state 
energy $E_s$ in Eqn(10) for different parameter values $t^\prime/t$, 
$U/t$ and $V/t$. For $t^\prime = 0$, we find that the CDW and SDW 
states are the only stable states and are separated by a phase 
boundary at $U=4V$. In Fig.2 we present the phase diagram obtained 
from Eqns(9) and (10) for typical values of $t^\prime/t$ in the 
($U/t,\,V/t$) plane. Our results indicate that the destruction of 
nesting, introduced by $t^\prime$ helps to stabilize the OAF/SN 
states at weak coupling. The various order parameters do not vanish 
at the phase boundaries which therefore indicate discontinuous phase 
transitions. The phase boundaries for small values of the interaction 
parameters $U/t$ and $V/t$ have not been shown as numerical 
difficulties prevented an accurate determination of their positions. 

  The zero temperature instability of the Fermi system
  to the states considered by us can be roughly understood 
  as arising from the energy gained due to the
  opening up of a gap $\Delta({\bf k})$ in the energy spectrum
  which pushes down (in energy) the occupied states.
  It is then easy to see from Eqn. (6) 
  that the gap functions in the CDW, SDW and OAF/SN states
  are proportional to  $4V-U/2$, $U/2$ and $2V$ respectively.
  Thus the OAF/SN states are expected to be energetically
  close to the other two states near the $U=4V$ line.
  At large values of U, the dominant consideration is to minimize
  the double occupancy in the ground state which
  is achieved by the SDW state with $m\rightarrow 1$.
  On the other hand for large values of $V$, the dominant tendency
  is for all the electrons to sit on one of the sublattices
  thus completely avoiding the energy cost associated with
  having electrons on neighbouring sites.
  This corresponds to the CDW state with $p\rightarrow 1$.
  Thus the OAF/SN states are expected to arise only
  at weak coupling and close to the $U=4V$ line. This
  is borne out by our numerical results.
  The crucial role of $t^{\prime}$ in stabilizing
  the  OAF/SN states is apparent from our phase diagram.
   The physical basis for this is not clear to
  us. However it suggests that the OAF/SN states are
  more robust with respect to the destruction
  of the nesting of the Fermi surface than the CDW
  or the SDW states.

In Fig.3 we show the dispersion $e_\vk^{(-)}$ and $e_\vk^{(+)}$ for 
the OAF (SN) states as a function of $\vk$ for $t^\prime/t$ finite 
with $U/t$ and $V/t$ chosen to ensure the stability of the phases 
considered. It is clear from Fig.3 that for $t^\prime \neq 0$ the 
upper and lower bands overlap.
Thus the valence band is not completely full and the conduction band
is not completely empty.
The OAF (SN) state is therefore a poor 
metallic state with a pseudogap at the Fermi surface.

Finally, we summarize the main results of this paper. We have 
obtained the zero temperature phase diagram of the 
extended Hubbard model on a square lattice at half-filling. We 
find that for finite values of next-nearest-neighbour hopping amplitude
$t^\prime$, the OAF/SN states 
have lower ground state energy than the SDW and CDW state over 
a finite range of parameters in the weak coupling region. The size 
of this region, where OAF/SN states are stable, increases with 
increasing $t^\prime$. 

We would like to thank A. Taraphder, R. Pandit and T. V. Ramakrishnan 
for useful discussions.

\newpage 

\noindent {\bf FIGURE CAPTIONS}

\noindent Fig.1. Schematic diagram of the circulating currents 
  associated with the OAF and SN states. The arrows indicate the 
  directions of the currents which are staggered and move oppositely 
  on neighbouring plaquettes. 

\vspace{0.2in}
\noindent Fig.2. The mean-field phase diagram of the $t-t^\prime$ 
  extended Hubbard model. All the phase boundaries indicate discontinuous 
  transitions. The region of stability of the OAF (SN) state grows with 
  increasing $t^\prime/t$ ratio, as seen in the figure. 

\vspace{0.2in}
\noindent Fig.3. Plot of the quasiparticle dispersion energy 
  $e_\vk^{(+)}$ and $e_\vk^{(-)}$ for the OAF and SN states along the 
  symmetry directions in the reduced Brillouin zone. The solid curve 
  indicates the conduction band dispersion while the dashed curve 
  is for the valence band dispersion. The parameters choosen for the 
  quasiparticle dispersion are $t^\prime/t=0.1$, $U/t = 1.5$ and 
  $V/t = 0.37$ for which the OAF (SN) state is stable. 
  Inset shows the position of the points $\Gamma$, $X$ and $\Sigma$
  in the Brillouin zone.


\newpage

\begin{thebibliography}{999} 

\bibitem{schr}{J. R. Schrieffer, X. G. Wen and S. C. Zhang, Phys. Rev. B 
  {\bf 39}, 11663 (1989).}
\bibitem{halp}{B. I. Halperin and T. M. Rice, Solid State Physics, Vol.  
 {\bf 21}, ed. F. Seitz, D. Turnbull and H. Ehrenreich, Academic Press, 
 New York (1968).}
\bibitem{shulz}{H. J. Shulz, Phys. Rev. B {\bf 39}, 2940 (1989).}
\bibitem{ner1}{A. A. Nersesyan and G. E. Vachnadze, J. Low Temp. Phys.  
  {\bf 77}, 293 (1989).}
\bibitem{ner2}{A. A. Nersesyan, G. I. Japaridze and I. G. Kimeridze, 
  J. Phys. Cond. Matt. {\bf 3}, 3353 (1991).}
\bibitem{gorkov}{L. P. Gorkov and A. Sokol, Phys. Rev. Lett. {\bf 69}, 
  2586 (1992).}
\bibitem{nayak}{C. Nayak and F. Wilczek, Preprint, cond-mat/9510132}
\bibitem{affleck}{I. Affleck and B. Marston, Phys. Rev. B {\bf 37}, 
  3774 (1988).}
\bibitem{marsh}{D. S. Marshall {\it et. al.}, Phys. Rev. Lett. {\bf 76}, 
  4841 (1996); A. G. Loesen {\it et. al.}, Science 273, 325 (1996).}
\bibitem{ding}{H. Ding {\it et. al}, Nature {\bf 382}, 51 (1996); Preprint, 
  cond-mat/9611194.}
\bibitem{emery}{V. J. Emery and S. A. Kivelson, Nature {\bf 374}, 
  434 (1995).} 
\bibitem{laugh}{R. B. Laughlin, Preprint, supr-con/9608005.} 
\bibitem{mermin}{N. D. Mermin, J. of Phys. Soc. Japan {\bf 26},
Supplement, 203 (1969)}


\end{thebibliography}
\end{document}